\AtBeginDocument{%
  \providecommand\BibTeX{{%
    \normalfont B\kern-0.5em{\scshape i\kern-0.25em b}\kern-0.8em\TeX}}}

\newif\ifisanon
\isanonfalse

\documentclass[10pt, conference, letterpaper]{IEEEtran}
\usepackage{subfigure}
\usepackage{xspace}

\usepackage{xcolor}

\usepackage{multirow}
\usepackage{url}
\usepackage{graphicx} 
\usepackage{hyperref}

\hypersetup{colorlinks,linkcolor=[rgb]{.75,0,0},urlcolor=[rgb]{0.75,0,0.75},citecolor=blue,breaklinks} 

\makeatletter
  \newcommand\EatSpacesHack{\@bsphack\@esphack}
\makeatother
\iftrue
  \newcommand\comment[1]{{\color{blue} \sffamily [xxx:  #1]}}
  \newcommand\reviewfix[1]{{\sffamily [RF:#1]}\@bsphack\@esphack}
  \newcommand\PostSubmission[1]{{\sffamily [post-xxx:  #1]}}
  \newcommand{\jelena}[1]{{\textcolor{purple}{jelena: #1}}}
  \newcommand{\rizvi}[1]{{\textcolor{brown}{rizvi: #1}}}
  \newcommand{\johnh}[1]{{\textcolor{blue}{john: #1}}}
  \newcommand{\wes}[1]{{\textcolor{magentare}{wes: #1}}}
  \newcommand{\robert}[1]{{\textcolor{green}{robert: #1}}}
\else
  \newcommand\comment[1]{}
  \newcommand\PostSubmission[1]{\EatSpacesHack}
  \newcommand\reviewfix[1]{\EatSpacesHack}
  \newcommand{\jelena}[1]{\EatSpacesHack}
  \newcommand{\rizvi}[1]{\EatSpacesHack}
  \newcommand{\johnh}[1]{\EatSpacesHack}
  \newcommand{\wes}[1]{\EatSpacesHack}
  \newcommand{\robert}[1]{\EatSpacesHack}
\fi

\def\Snospace~{\S{}} %
\newcommand{\system}{\emph{DDiDD}\xspace}
\ifisanon
        \newcommand{\broot}{\emph{$root_{ANON}$}}
\else
        \newcommand{\broot}{\emph{B-root}}

\fi
\makeatletter
  \def\section{\@startsection{section}{1}{\z@}%
  {1.0ex plus 1.0ex minus 0.5ex}%
  {0.5ex plus 1ex minus 0ex}%
  {\normalfont\normalsize\centering\scshape}}%
  \def\subsection{\@startsection{subsection}{2}{\z@}%
  {0.8ex plus 0.6ex minus 0.4ex}%
  {0.3ex plus .3ex minus 0ex}%
  {\normalfont\normalsize\itshape}}%
\makeatother

\def\BibTeX{{\rm B\kern-.05em{\sc i\kern-.025em b}\kern-.08em
    T\kern-.1667em\lower.7ex\hbox{E}\kern-.125emX}}

\begin{document}
\title{Defending Root DNS Servers Against DDoS \\Using Layered Defenses}
%
%
\author{\IEEEauthorblockN{A S M Rizvi,
Jelena Mirkovic, John Heidemann, Wesley Hardaker and
Robert Story}
\IEEEauthorblockA{University of Southern California / Information Sciences Institute}}

%
%
%
\maketitle              
\begin{abstract}
Distributed Denial-of-Service (DDoS) attacks exhaust resources,
  leaving a server unavailable to legitimate clients.
The Domain Name System (DNS) is a frequent target of DDoS attacks.
Since DNS is a critical infrastructure service, protecting it from DoS is imperative.
Many prior approaches have focused on specific filters or anti-spoofing techniques
  to protect generic services. 
  DNS root nameservers are more challenging to protect, since they use fixed IP addresses, serve very diverse clients and requests, receive predominantly UDP traffic that can be spoofed, and must guarantee high quality of service.
In this paper we propose a layered DDoS defense for DNS root nameservers.
Our defense uses a \emph{library} of defensive filters,
  which can be optimized for different attack types,
  with different levels of selectivity.
We further propose a method that \emph{automatically and continuously evaluates and selects} the best combination of filters throughout the attack. 
We show that this layered defense approach provides exceptional protection against all attack types using traces of ten real attacks from a DNS root nameserver. 
Our automated system can select the best defense within seconds and quickly reduces traffic to the server within a manageable range, while keeping collateral damage lower than 2\%. We can handle millions of filtering rules without noticeable operational overhead. 
\end{abstract}
%
%
%
\section{Introduction}

 Distributed-Denial-of-Service (DDoS) attacks
  remain a serious problem~\cite{akamai-report,anstee2013worldwide,microsoft2021,akamai2021}, in spite
  of decades of research and commercial efforts to curb them. 
  Ongoing Covid-19 pandemic and increased reliance of our society
  on network services, have further increased opportunities for DDoS attacks. 
According to the security company F5 Labs,
  between January 2020 and March 2021,
  DDoS attacks have increased by 55\%~\cite{f5labs}. 
While some large-volume DDoS attacks make front page news
  (for example, the 1.35\,Tb/s~\cite{memcache} attack on Github in Feb.~2018,
  or 2021 17.2\,M requests per second attack, detected by CloudFlare~\cite{cloudflare2021}),
 many more attacks occur daily and disrupt
operations of thousands of targets~\cite{dyn-ddos,ddos2021voipms}.


This paper focuses on protecting
  the Domain Name System (DNS) root servers against DDoS attacks.
The root-DNS service is a high-profile, critical service,
  and it has been subject to repeated
  DDoS attacks in the past~\cite{Vixie02a,root-news-nov,root-news-june,moura2016anycast,dyn-1}.
In addition, because the DNS root ``bootstraps'' DNS,
  it is served on specific IP addresses that cannot be easily modified,
  thus precluding use of many traditional DDoS defenses that redirect traffic to clouds to distribute load~\cite{Calder15a}.
  
There are many types of DDoS attacks.
Some attacks are conceptually easy to mitigate with firewalls, assuming upstream capacity is sufficient, such as volumetric attacks using junk traffic. 
Others, such as exploit-based attacks,
  remain pernicious, but automated patching and safer coding practices  offer promise.
Most challenging are attacks using legitimate-seeming application traffic,
  since a \emph{flash-crowd attack} from millions of compromised hosts (also known as layer-7 or application-layer attacks) can resemble a legitimate \emph{flash crowd}, when many legitimate clients access popular content. 
At DNS root servers, flash crowd attacks would generate excessive DNS queries.
Because legitimate clients also generate DNS queries, it is challenging to filter out attack traffic.
We focus on mitigation of flash-crowd attacks on DNS root servers. 

In flash-crowd attacks,
  attack traffic often appears identical in content to legitimate traffic.
Approaches to handle flash-crowd attacks thus focus on withstanding the attack using cloud-based services~\cite{Dilley02a,Pang04a,Moura18b,Rizvi22a}. Other approaches aim to separate legitimate from attack clients, e.g., via CAPTCHAs ~\cite{oikonomou2009modeling}, or by using models of typical client behavior~\cite{ramanathan2020blag,tandon2020quantifying}. 
These defenses work poorly for DNS root servers.
First, the DNS root operates at small number of fixed IP addresses
  that cannot be easily changed.
This restriction precludes use of  
  traditional defenses that redirect traffic to clouds~\cite{Calder15a}.
Second, DNS traffic to roots is generated by recursive resolvers.
Since there is neither direct interaction with a human
  nor a web-based user interface, CAPTCHAs cannot be interposed.
Third, aggressive client identification
  requires modeling a typical legitimate client.
Building a typical client model at roots is challenging, because
 client request rates vary by five orders of magnitude, from a few queries per day to thousands of queries per second.
A model
  that spans all types of clients can be too permissive, while a model that captures a majority of clients may drop legitimate traffic from large senders. 
Since most DNS traffic is currently UDP-based, spoofing also is a challenge
  and spoofers can masquerade as legitimate clients.

In this paper we propose a multi-layer approach to DNS root server defense against DDoS attacks, called \system ~-- DDoS Defense in Depth for DNS.
Our first contribution is to \emph{propose an automated approach to select the best combination of filters for a given attack}.
Selecting from a library of possible filters is important,
  since different filters are effective against different attacks,
  and each filter has a different false positive rate, and different operational cost, which precludes its continuous use. 
\system selects the best combination of filters quickly (within 3\,s)
  and continuously re-evaluates filtering effectiveness. When attack traffic changes (e.g., in case of polymorphic attacks), \system quickly detects decrease in the filtering effectiveness and re-selects a new, better combination, thereby adjusting to dynamic attacks.

Our second contribution is
 to propose a novel \emph{wild client filter for DNS}.
We provide the first open description and evaluation of a filter
  that models per-client behavior for DNS clients. 
Client modeling is widely used to protect web servers~\cite{tandon2021defending}
  where a single model for a ``typical'' web client suffices.
DNS shows a huge
range of rates (over 5 orders of magnitude)
  across clients, so any model that captures this entire range will be too permissive. 
Instead, we model each client separately during pre-attack periods, and identify as attackers the clients that become more aggressive during attacks. In deployment we combine this filter with anti-spoofing filters to establish trust in client identities.

Our final contribution is to perform \textit{evaluation of each candidate filter}, including our wild resolver filter and six other filters proposed in prior work~\cite{schomp2020akamai, wang2007defense,jin2003hop,mukaddam}. While prior work quantified performance of some individual  filters for general DDoS attacks~\cite{wang2007defense, jin2003hop, mukaddam},
and other work qualitatively described commercial deployments (such as Akamai's~\cite{schomp2020akamai}),
  we are the first to  evaluate each filter quantitatively against real DDoS attacks on a DNS root.
We are also the first to propose and evaluate a dynamic multi-filter system for protection of DNS roots against DDoS.
Our evaluation uses \emph{real-world attacks and normal traffic taken
  over 6 years from \broot},
  as well as an \emph{adversarial, polymorphic attack} we have synthesized.
Our evaluation confirms that no
  single filter outperforms the others,
  but together they provide a stable defense against different attack types
  converging in 3\,s or less, with low collateral damage (at most 2\%).
Our analysis provides evidence for the DNS operators
  about the importance of having an automated system, and 
  it provides insights about individual filter performance against different types of attacks.



We focus our work on the DNS root server system
  to meet its unique challenges,
  but 
  our results also apply to other self-hosted, authoritative DNS servers.

We release the DDoS datasets that we use in this paper\ifisanon~\cite{dsnanonAnon}\else~\cite{attackdatasets}\fi.

\iffalse
%
%
\section{Background: DNS and DDoS}

Domain Name System (DNS) is part of the critical Internet infrastructure. It maps between resource names and IP addresses, using a hierarchical database distributed across \textit{authoritative DNS nameservers} (\emph{authoritative} for short).
The DNS root is on top of the hierarchy, followed by top-level domain (TLD) servers and subdomain servers. Each authoritative nameserver is responsible for maintaining mapping of some portion of the DNS namespace, and for replying to queries about that portion to any DNS client.
In addition to authoritative nameservers, there are also \textit{recursive resolvers} (\textit{recursives} for short), which receive queries from end-user devices, and then query one or more authoritative servers to resolve the query, and send the response back to the end user.  There are many recursives, and each serves some local users by proxying for them the translation of DNS names to IP addresses, and caching any new data learned from the authoritative servers. From an authoritative server's point of view, these recursives represent \textit{DNS clients} that interact with this server.

Each DNS name consists of multiple components, separated by periods, such as \texttt{www.example.com}.
The rightmost segment denotes a top-level domain or TLD, such as \textit{.com} or \textit{.us}.
When a recursive resolver looks up a name, it parses each component,
  querying authoritative nameservers, if it does not have a resolution for that given suffix in its cache.
Components and their resolutions are cached for durations specified by
  their owner, and can be overridden by the recursive's configuration.

A recursive resolver bootstraps using a pre-configured list of DNS root servers.
It refreshes this list with a \emph{priming query}~\cite{Koch17a}
  to any server on the list. Updates to the list occur rarely, on timescales of months or years,
  due to careful planning and the software upgrade cycle.
Domains served by the root (top-level domains, or TLDs)
  are typically cached by recursives for 24 or 48 hours.

\subsection{DNS Root Traffic}

Because resolvers cache responses from DNS root, and because TLD records have large TTL values and there are only a few thousand TLDs, one would expect that each resolver queries a root server very infrequently. The actual traffic from resolvers, however, defies this expectation by a large margin. 
Figure \ref{fig:roottraffic} illustrates the complementary cumulative distribution function (ccdf) of the number of queries to \broot{} per hour at a random hour in each of the years 2015, 2016, 2017, 2018 and 2019. There is a wide range of rates across 5 degrees of magnitude.
While majority of resolvers exhibit the behavior we expect -- 95\% send fewer than 62 queries per hour and 99\% send fewer than 1,500 queries per hour -- a small number of resolvers sends excessive numbers of queries, up to 100,000 queries per hour!

\begin{figure}
\begin{center}
\includegraphics[width=3in]{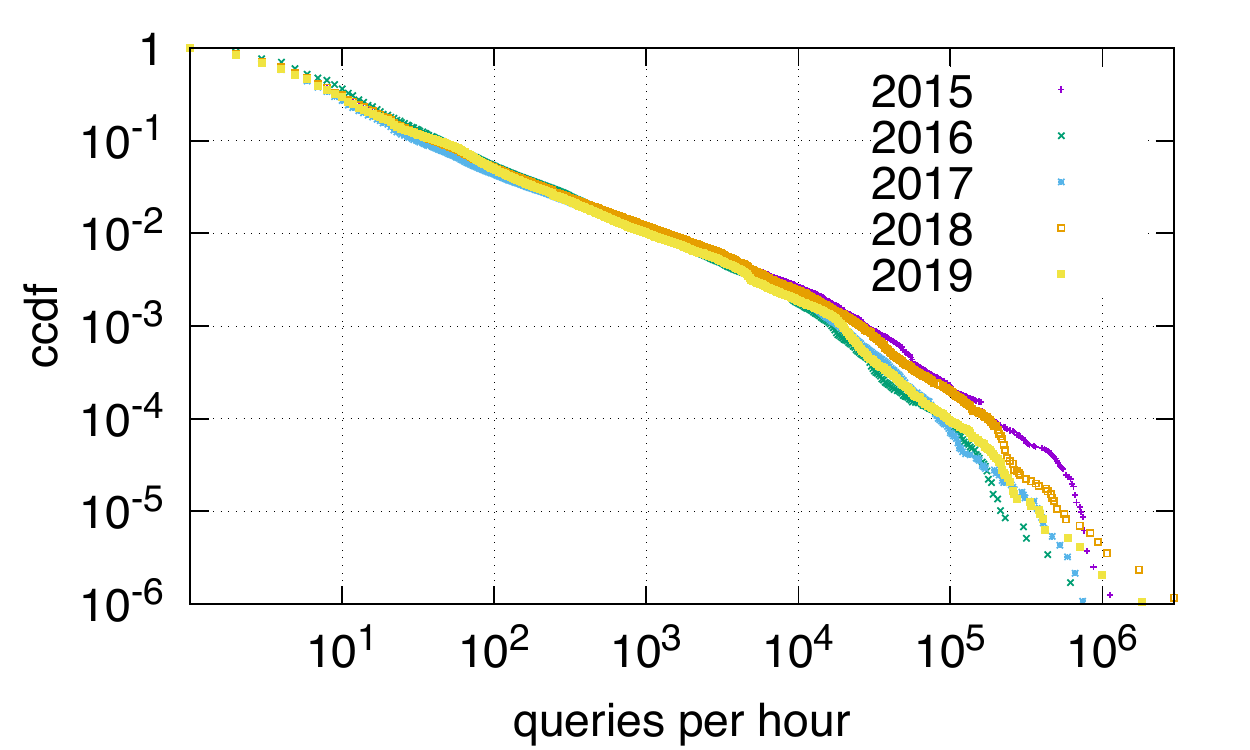}
\caption{Complementary cumulative distribution function of the number of requests per hour sent to \broot{} on five random dates between 2015 and 2019. }
\label{fig:roottraffic}
\end{center}
\vspace{-0.20in}
\end{figure}

Why are roots receiving so much more traffic than one would expect? There are several possible explanations. First, roots receive many queries (36--39\% in our dataset) that do not have a valid TLD~\cite{a-root-rcode, b-root-rcode, castro2008day}. Chrome browsers make such random DNS queries to detect DNS hijacking~\cite{dns-chrome}. Roots will return a no-such-domain (NXDomain) reply to these queries, but such replies are not cached by resolvers.
\comment{this next sen doesn't make sense.  by definition they are resolvers because they are making dns queries.  I think the point is: ``some may not be serving users, but instead may be monitoring the root.'' ?
(I dont' think we should hint at ``some unknown purpose''. ---johnh 2022-02-22}
Second, some resolvers may not be caching properly, and may ask for the same, existing TLDs, repeatedly. Third, some resolvers may not actually be serving end users, but instead they may be programmed to query the root by some application.\jelena{I am not sure that they are monitoring the root. we simply don't know why they behave the way they do. This is what I was trying to say.}
More research is needed to establish root causes of this excessive traffic.

Root servers operate as a service to the Internet and are committed to
  serving the root DNS zone as defined by IANA
  to all queriers
  (for example, see~\cite{BRoot08a}).
\rizvi{We can just cite it without saying ``for example``.---2022-02-27}
Due to this policy, root server operators prioritize responding to all queries,
  with the exception of obvious attacks and operational threats.

\subsection{The DNS Root and DDoS}

Historically there have been several large attacks on DNS root servers.
In 2002~\cite{attack2002}, a large volumetric attack hit all 13 DNS root servers for an hour, with nine of 13 root servers largely inaccessible.
In 2007~\cite{attack2007}, a volumetric attack hit six DNS root servers, and lasted 3\,h and 5\,h.  Two servers were noticeably affected.
In November/December 2015~\cite{moura2016anycast}, most of the root name servers were hit by two volumetric attacks containing millions of spoofed queries per second.
While some root servers were lightly hit, others saw severe traffic loss of 95\% or more.
Analysis showed the attacks inflicted collateral damage to services collocated with root servers~\cite{moura2016anycast}.
We use data from these attacks, and from eight attacks in the years following 2015.

\comment{I'm dropping this paragraph, it contridicates prior published results.
\cite{moura2016anycast} saw NO user-visible problems (see section 2.3).
And the recent 2021 sigcomm paper by Koch et al.~is an entire paper showing with root problems have minimal user impact.
---johnh 2022-02-23}
\jelena{let's chat about this then. we need to establish why handling DDoS attacks on DNS matters, when prior attacks were not devastating}
\comment{old text to drop: ``Even though attacks on root servers are not that frequent, a successful DDoS attack on one or more DNS roots can have a devastating impact on the root's clients. A client that receives no response will send its query to another root server, but the name resolution delay will be visible by the end-user. In addition to this, some DNS clients do not switch to other root servers and thus experience sustained damage~\cite{moura2016anycast}. Additional queries can also overwhelm other root servers, slowing down their processing. Since resolving DNS is the first step for most web interactions, clients experience higher latency. In the worst case, clients observe service interruption because of the failure to resolve DNS~\cite{facebook-dns}.''}

\comment{(1) I rewrote this paragraph next,
but... (2) we need to make the case for novelty in the intro, not here.  here is WAY too late.
(3) so can we remove it? ---johnh 2022-02-23}
\comment{old para: 
``While DDoS attacks are an old, and well researched phenomenon, DDoS on DNS root servers poses some unique challenges: (1) \textbf{IP spoofing.} DNS queries are prevalently sent over UDP, and thus can be spoofed. This complicates defenses that aim to identify and block aggressive clients during attacks. (2) \textbf{Query rate diversity.} DNS resolver query rates span five orders of magnitude. This complicates defenses that model a client's sending rate during no-attack periods. Individual attackers can easily fit this model, while in aggregate overwhelming the root server. (3) \textbf{Requirement for low collateral damage.} DNS root servers share a commitment to serve every DNS query. Thus, filtering queries to mitigate a DDoS attack is the action of the last resort. Filters must be very accurate, and minimize collateral damage.''
}

While DDoS attacks have been studied for decades,
  DDoS on DNS root servers pose some unique challenges.
Unlike web traffic, most DNS queries use UDP and UDP support is required,
  so DNS is vulnerable to \emph{IP spoofing},
  making filtering by source IP address ineffective.
Second, root severs see a huge \emph{diversity of query rates}---legitimate traffic
  spans five orders of magnitude, complicating traffic modeling
  and filtering big senders.
Third, \emph{root DNS} uses a small number of fixed IP addresses,
  so shifting traffic to other anycast rings is not feasible.
Finally, the DNS root has a very high commitment to serving all queries,
  so \emph{collateral damage} is a large concern;
  good queries should be answered.
  
\else

\section{Background: DNS and DDoS}

The Domain Name System (DNS) is critical Internet infrastructure
  that maps between human-readable names and resources such as IP addresses.
DNS names are hierarchical,
  with the root, top-level domains (TLDs), like \texttt{.com} and \texttt{.uk},
  and subdomains, like \texttt{example.com}.
This hierarchy is distributed across many
  \emph{authoritative nameservers} (``authoritatives'' for short).
Users usually do not directly query the DNS,
  but instead use \emph{recursive resolvers} (``recursives'' for short)
  that resolve names on their behalf.
Each recursive usually provides service for many users, caching responses to speed access. 

For resilience, root zone is served by 13 identifiers,
  each at a unique, anycasted IPv4 and IPv6 address, served by multiple authoritative servers at multiple geographical points of presence (PoPs). Three aspects make  the authoritatives for the DNS root challenging to defend from flash-crowd DDoS attacks. 
First,
  most DNS queries use connectionless UDP (not TCP),
  so it is trivial for an attacker
  to spoof source IP addresses, making defenses that model client behavior unreliable.
Second, root authoritative servers see a huge range of query rates from different recursives---over five orders of magnitude, and huge query content diversity.
This variation makes it impossible to produce a single, tight model for a ``typical recursive behavior''. 
Third, the DNS root is used to bootstrap the DNS system,
  and so it operates at fixed IP addresses.
Although resolvers refresh this list on startup~\cite{Koch17a}, the list is expected to be mostly static. Deploying new root servers takes months of careful planning. Thus defenses typically used by Content Delivery Networks (CDNs) to shift traffic to different servers (such as~\cite{Calder15a}) cannot be used to protect DNS root.

%
%
Because of its visibility and defensive challenges,
  the DNS root has been the target of several DDoS attacks.
During large, volumetric attacks in 2002~\cite{attack2002},
  2007~\cite{attack2007},
  and 2015~\cite{moura2016anycast},
  several of the 13 root identifiers showed service degradation (we show other events in \autoref{sec:eval}).
Although caching of root contents at recursives reduces the end-user impact of these attacks~\cite{moura2019cache,Koch21a},
  DNS outages at CDNs have impacted prominent user-facing services~\cite{dyn-ddos}.
Effective DDoS defense for the DNS root is thus necessary.

\fi

\section {Related Work} 
	\label{sec:related}

DDoS attacks have been a problem for more than two decades,
  and many research and commercial defenses have been proposed.
This section reviews only those solutions that 
  are closely related to our approaches and to protecting DNS servers against DDoS. 


\subsection{Flash-Crowd DDoS Defenses}

CAPTCHAs~\cite{Barna,kandula2005botz} are a popular defense against flash-crowd attacks. They can be used together with other indicators of human user presence, to differentiate between humans and bots. However,
DNS queries come from recursives, not directly from human users,
  so there is no opportunity for a CAPTCHAs intervention.
FRADE~\cite{tandon2021defending} is a flash-crowd DDoS defense, which builds models of how human users interact with a Web server, including query rates and query content, and uses them to detect bot-generated traffic. 
FRADE models a \textit{typical client's} behavior. While this works for Web servers, which are browsed by humans, request rates and contents of DNS recursives vary widely.
FRADE thus cannot protect DNS servers against DDoS.

Creating an allow-list of known-good clients is  
  suggested in several studies and
  RFCs~\cite{chen2010whitelist,yoon2010using,peng2004proactively, senie1998network,levine2010rfc} for general protection from unwanted traffic. However, the approaches to create a list of known-good recursives for DNS roots have not been described nor evaluated.
We evaluate this idea in this paper under the name ``unknown recursive filter,'' in conjunction with hop-count filtering~\cite{jin2003hop}, and show that it works well to filter out spoofed attack traffic, but cannot handle attacks that do not use IP spoofing.

Many companies provide DDoS solutions, which may combine signature-based filtering, rate limiting, and traffic distribution using cloud resources and anycast.
Such solutions are offered by Akamai~\cite{schomp2020akamai,gillman2015protecting}, Verizon~\cite{verizon-1}, and
  Cloudflare~\cite{cloudflare_ddos_1,cloudflare_ddos}, for example. 
  Since these solutions are proprietary, we cannot compare against them directly.
In addition,
  they often collect traffic with
  DNS-based redirection or route announcement (friendly hijacking).
Neither of these redirections are possible for root DNS service,
  which must operate at a fixed IP address, 
  and cannot easily be re-routed.

\subsection{Spoofed Traffic Filtering}

Several filters to remove spoofed traffic have been proposed:
  hop-count filtering~\cite{wang2007defense,jin2003hop,mukaddam}, 
  route traceback~\cite{sung2003ip}, route-based filtering~\cite{duan2008controlling}, path identifier~\cite{yaar2006stackpi},
  unknown client filtering~\cite{yoon2010using,peng2004proactively},
  and client legitimacy based on network, transport and application layer   information~\cite{thomas2003netbouncer}.
Of these approaches, only hop-count filtering and unknown client filtering can be deployed on or close to the target, and thus show promise for protection of DNS root servers. 
In hop-count filtering, the filter learns which IP TTL values are used in packets from a given source IP address, and uses this to filter out spoofed packets. 
The original approach~\cite{wang2007defense} advocates for storing one expected hop-count per source.
Mukaddam et al.~show that recording a list of possible hop-counts improves the precision of TTL filters~\cite{mukaddam}. These studies are performed on 
10--20 years old traceroute measurements, and they assume reliable inference of TTL filters from established TCP connections. Both Internet topology and application dynamics have since evolved, and DNS traffic is predominantly UDP. Our paper fills this gap, by evaluating hop-count filtering 
against DDoS with real attack and legitimate traffic, spanning six years and ten attack events.


\subsection{DDoS on DNS}

BIND pioneered Response Rate Limiting (RRL) to avoid excessive replies ~\cite{isc-rate} and conserve outgoing network capacity during a volumetric query DDoS.
RRL addresses a few misbehaving clients and outgoing amplification attacks,
  but it does not address well-distributed, volumetric attacks from large botnets.

Akamai uses sophisticated scoring and priority queuing to protect their authoritative DNS servers from floods~\cite{gillman2015protecting, schomp2020akamai}.
Akamai scores queries with the source's expected rate,
  if the resolver participated in prior attacks,
  the source's NXDomain fraction,
  query similarity from that source,
  and an evaluation of TTL consistency.
While two of these scoring approaches are similar to our unknown resolver and wild resolver filters, there are three major differences.
First, 
  Akamai provides no quantitative data about how various scoring approaches perform against real attack events. 
We contribute 
  a careful \textit{quantitative evaluation} of how well different filters work
  against playback of real attacks.
Second, we propose a specific mechanism to select filter combinations, and reevaluate them when necessary. Akamai's approach uses all filters at once to calculate each query score, and Schomp et al.~\cite{schomp2020akamai} does not describe how the filters interact.
Finally,
  key parts of Akamai's scoring system run inline with processing,
  requiring high-speed packet handling.
Our approach operates in parallel with packet processing,
  evaluating resolvers to identify potential attackers (or known-good resolvers),
  simplifying deployment, particularly for lower-end hardware.

  
Prior work has studied real DDoS events,
  inferring operator responses using anycast,
  and suggesting possible anycast options in DNS roots~\cite{moura2016anycast}.
Recent work has taken this idea further,
  suggesting that a network playbook
  can pre-evaluate routing options to shift traffic across anycast sites~\cite{Rizvi22a}.
Our work complements this line of research,
  by studying how filters can reduce load at each anycast site.

Finally, several groups have suggested fully distributing the root
  to all recursives~\cite{wjh:ndss2018dnsprivacy,Allman19a,Kumari20a}.
Such wide replication would greatly reduce the threat of DDoS on the root,
  but not on other DNS authoritative servers.
As a result,
  on-site defense is still necessary to mitigate DDoS attacks on DNS.

\section{\system Design}
	\label{sec:design} 

Our goal is to design an automated system, which continuously 
evaluates suitability of multiple filters to handle an ongoing DDoS attack on a DNS root server.  
Our system needs to quickly select the best filter or the combination of filters, reasoning about the projected impact on the attack, the collateral damage from the filter on legitimate recursives' traffic and the operational cost. The system should also be able to adjust its selection as attack changes. Finally, individual filters need to be configured to achieve optimal performance -- high effectiveness against attacks they are designed to handle and low collateral damage.

DNS root may also experience a legitimate flash crowd, e.g., when many clients access some popular online content. Due to caching, queries for existing TLDs should not create flash crowd effect, but queries for non-existing TLDs may, since their replies are not cached. \system will only activate when excessive queries overwhelm server resources. Unless the server can quickly draft more resources (e.g., through anycast) some queries have to be dropped. Without \system, random legitimate queries would be dropped. \system (\autoref{sec:eval}) mostly drops queries from sources causing the legitimate flash crowd. 

\subsection{Threat Model}
  	\label{sec:back_ex}

We assume that an attacker's goal is to exhaust some key resource at a target by sending legitimate-like requests to the server.
Current authoritative servers (including root) do not store
  state between requests,
  so the attacker can target CPU resources, incoming bandwidth or outgoing bandwidth.
In all cases, the attacker generates more requests than the server can process per second.
The attacker may spoof these requests, or they may compromise new or rent existing bots and send non-spoofed requests. 

A \textit{spoofing attacker} may spoof at random, or they may choose specific IP addresses to spoof.  In some cases, the attacker may choose to spoof addresses of existing, legitimate recursives.

\begin{figure*}[t!]
\begin{center}
                 {\label{fig:pseudocode}
         \includegraphics[width=0.7\textwidth]{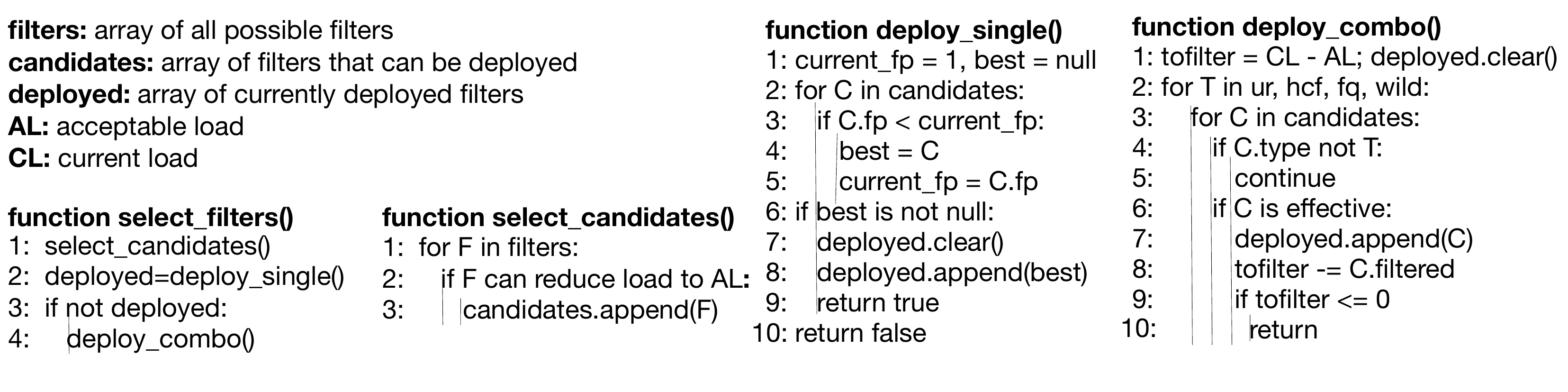}}
\caption{Pseudocode for filter selection}
\label{fig:pseudocode}
\end{center}
\vspace{-0.27in}
\end{figure*}

A \textit{non-spoofing attacker} compromises or rents bots to use in the attack.
Drafting new bots carries non-negligible cost for the attacker.

The features of \textit{attack requests} depend on the resource that the attack targets. If the targeted resource is CPU, the attacker may generate many requests per second. If the target is incoming bandwidth, the attacker may generate large requests to quickly consume the bandwidth. In both of these cases, the content of the requests is not important, just their rate and size. Finally, if the target is outgoing bandwidth, the attacker may generate requests that maximize the size of replies, using the ANY query type. 

Some attacks are \emph{polymorphic} -- they change their features during the attack event.
Any attack features may change: how spoofing is done, which sources generate attacks,
  and the content of attack requests.

A \textit{naive} attacker does not have knowledge about \system and is focused only on overwhelming the target server. A \textit{sophisticated} attacker may obtain information about types and parameters of the filters that our defense uses, and they may try to adjust their attack to bypass the defense, or to trick the defense into filtering a legitimate recursive's traffic.

\system works well both against naive and against sophisticated attackers, and against spoofing and non-spoofing attackers, due to its layered defense approach, and multiple filters, as we show in our evaluation.

\subsection{\system Operation}
  	\label{sec:operation}

To avoid any operational impact on a DNS root server,
  \system consumes packet captures,
  operating offline, independently of the actual DNS server software.
\system's analysis detects an attack,
  selects a filter or a combination of filters, then deploys filters via
  \texttt{iptables} and \texttt{ipset} rules on the server. We consider six filters, described in \autoref{sec:filters}, and implement four that perform well with DNS root traffic: frequent query filter, unknown recursive, wild recursive and hop-count filter. 
\texttt{iptables} work well when number of rules is small (up to 2\% delay increase for 5 rules) and matching is needed on query content. We use \texttt{iptables} to implement the frequent query filter, for 1--5 frequent query names. \texttt{ipset} uses an indexed data structure
  and provides efficient matching of thousands or even millions of rules, without added delay. We use it when blocking attack sources, identified by unknown recursive, wild recursive and hop-count filters. 
  \texttt{iptables}/\texttt{ipset} or their equivalents are available on all modern operating systems, thus \system is highly deployable by any interested DNS root server. If a root is anycast over multiple points-of-presence (PoPs), \system should be deployed at each PoP independently. No synchronization or information exchange is required across instances deployed at different PoPs. 
  
\system automatically selects filters to meet two goals.
First, we prefer filters
  that will remove most attack traffic with low or zero collateral damage
 to legitimate queries.
Second, we aim to select filters quickly, because most DDoS attacks are short~\cite{caidamacroscopic}. We then revise our selection if attack changes, or if we learn that another filter combination works better. This decision process is fully automated.
Further, \system is flexible and modular, allowing addition of new filters in the future. 

\textbf{Attack detection.}
\system detects attacks by monitoring the status of critical resources
  and recognizing when a resource is overloaded.
We use \emph{collectd} to periodically collect status information
  from several resources (CPU, memory, inbound and outbound network capacity).
We identify attacks when any resource exceeds a   
  fraction of its maximum capacity, which we denote as \textit{critical load}. 

We detect attack termination
  by monitoring the amount of traffic blocked by the deployed filters. 
We declare the attack over when the traffic blocked by \system decreases significantly, and the load on the server stays low as well, for an extended period of time. 
More details are given in\ifisanon~\cite{dynamictranon}\else~\cite{Rizvi19a}\fi.


\textbf{Filter priming and selection. }
All filters (e.g., frequent query filter, unknown recursive, wild recursive filter, hop-count filter) require information that must be learned continuously, in absence of attacks. \system continuously learns these parameters from packet collection and uses them when the corresponding filter is deployed. Some filters (e.g., frequent query name) also require a short learning phase during an attack. \system triggers a short learning phase for these filters when the attack is detected, and repeats it regularly. 

During attack, each filter and some filter combinations are continuously evaluated for potential deployment.
We emulate the effect of each filter or their combination on a sample of captured packets.
We estimate the success of each filter
  based on acceptable query load at the server, calculated as the server's average query load times a small multiplicative factor $f_{ACC}$. Because root servers operate well below their capacity, this approach guarantees that query rates below the acceptable load will also not exhaust the server's CPU or bandwidth resources, and will not trigger attack detection.

We also estimate collateral damage when the filter is parameterized
  using peace-time (non-attack) traffic.
The collateral damage depends on the legitimate traffic's blend and we have verified that it does not change sharply over time. Thus, we can calculate it once and use this estimate for a long time (e.g, months). 
Based on the estimated effectiveness of the given filter or their combination, and their projected collateral damage, new filters may be selected for deployment and existing filters may be retired.

 
\begin{table} 
    \centering
        \scriptsize
    \begin{tabular}{c|c|c}
        \textbf{parameter} &\textbf{meaning} & \textbf{rec. values}  \\ \hline
$L_{FQ}$ & num. queries for learning & 10~K\\
$f_{FQ}$ & freq. change threshold & 0.3 \\
$L_{UR}, L_{HC}, L_{WR}$ & learn. period & 2~h (20~m for WR) \\
$U_{UR}, U_{HC}, U_{WR}$ & use period & 2~h \\
$w_i$, ..., $w_N$. & observ. windows & $2^0$, $2^1$, ..., $2^8$ \\
 $t_{WR}$ & deviance threshold & 0.5 \\ 
    \end{tabular}
    
    \caption{Filter parameters}
    \label{tab:params}

    \vspace{-0.27in}
\end{table}

\subsection{\system Filters}
	\label{sec:filters}

In \system we have implemented the following filters: 
(FQ) frequent query name filter, (UR) unknown recursive filter, (HC) hop-count filter and (WR) wild recursive filter. In addition to these, we have also considered (RC) response-code filter and (AR) aggressive recursive filter. Since these two filters do not perform well on root server traffic, we do not include them in \system, but we evaluate them on our dataset and summarize results in this section. We show our recommended filter parameters in \autoref{tab:params}.
For each filter, we measure the performance and operational cost.

\textbf{Frequent query name filter (FQ).}
	\label{sec:qname}
In our datasets many attacks have queries that follow a given pattern, e.g., have a common suffix. Thus, in practice it is useful to develop filters that remove frequent queries during attack periods.

  



\textit{Approach:}
We use a simple algorithm to identify frequent query names. We continuously observe $L_{FQ}$ queries of incoming traffic and learn frequency of top-level domains, subdomains and full queries. Under attack, we repeat the calculation and look for segments (TLDs, subdomains or full queries) whose frequency has increased more than a threshold $f_{FQ}$.
These segments are candidates for frequent query names. Segment frequency prior to the attack serves to estimate collateral damage. We evaluated a range of values for $L_{FQ}$ and $f_{FQ}$. Shorter $L_{FQ}$ than 10,000 reduced mitigation delay, but increased chances of mis-identification of frequent queries. Similarly, lower $f_{FQ}$ than 0.3 lead to some collateral damage. These values should be calibrated for each server.

\textit{Operational cost:} We can filter frequent query names directly using \texttt{iptables}, or we can identify sources that send frequent queries and block them using \texttt{ipset}. We denote these two implementation approaches as FQ${_t}$ and FQ${_s}$. The FQ${_t}$ (\texttt{iptables}) implementation imposes added processing delay, which greatly increases once we go past five filtering rules, but it minimizes collateral damage. The FQ${_s}$ (\texttt{ipset}) implementation adds no measurable delay, but it may create collateral damage if spoofing is present, and thus must be deployed together with anti-spoofing filters (UR and HC). 

\textbf{Unknown recursive filter (UR).}
\label{sec:source_ip} 
An allow-list with IP addresses of recursives present prior to the attack can be an effective measure against random-spoofing attacks or those that rent bots. This filter passes traffic from recursives on allow-list to the server, and drops all other traffic.


\textit{Approach:} An allow-list is built by processing incoming traffic to the DNS root server over period $L_{UR}$ prior to an attack event. The list is then ready to be used for some time $U_{UR}$, and after that it can be replaced by new list.

\system builds allow-lists proactively at all times, observing traffic over period $L_{UR}$.  We experimented with $L_{UR}$ ranging from 10 minutes (capture 93\% of traffic sources) to 6 hours (capture 99\% of traffic sources). 
We also tested values of $U_{UR}$ of up to 1 day, and the allow-lists were very stable. 

\textit{Operational cost:} An allow-list can be implemented efficiently using  \texttt{ipset}, which adds no processing delay.

\textbf{Hop count filter (HC).}
\label{sec:hcf}
A hop-count filter builds the \textit{TTL-table}, containing source IP addresses, along with one or more TTL values seen in the incoming traffic from each given source. 
This kind of filter can be effective for attacks that spoof IP addresses of existing recursives. The filter drops traffic from sources that exist in the TTL-table, but whose TTL value does not match the values in the table.  All other traffic is forwarded.

\textit{Approach:} We build the TTL-table by processing incoming traffic to the DNS root server over period $L_{HC}$.  The list is then ready to be used for some time $U_{HC}$, and after that it can be replaced by new list. 

One could use hop counts~\cite{wang2007defense,mukaddam} or TTL values for filtering. TTL values are better choice, since they have larger value space, which improves filter effectiveness.  
\system builds its TTL-list by using each packet in the incoming traffic to the server during the learning period.
Such traffic could be spoofed. Prior approaches~\cite{wang2007defense,mukaddam,ttldrdos} rely on established TCP connections or they probe sources to reliably learn TTL-table values. These approaches do not work for DNS root servers, which serve mostly UDP traffic and whose policy forbids generation of unsolicited traffic. Hop-count filter parameter values have similar properties to known-recursive parameter values.

\textit{Operational cost:} We implement this filter efficiently by adding a new \texttt{ipset} module to match on an IP address and TTL value (or range). 

\textbf{Wild recursive filter (WR).}
\label{sec:drop}
While query rate of different DNS recursives towards a DNS root server varies widely, 
individual recursives' behaviors are mostly consistent over short time periods (e.g., several hours). We leverage this observation to build models of each individual recursive's behavior. The model for a given recursive, along with the recursive's IP address is stored in the \textit{rate-table}. During an attack, we identify those recursives that send more aggressively than their rate-table predicts as \textit{wild recursives}. Wild recursive filter drops traffic from wild recursives, and it forwards all other traffic.

\textit{Approach:} 
A wild-recursive filter learns the rate of a DNS recursive's interaction with the DNS root server over multiple time windows, $w_1, w_2, w_3, ..., w_N$, during learning period $L_{WR}$. For each window, the filter learns the mean and standard deviation of the number of queries observed and stores them in the rate-table. The rate-table can be used for some time $U_{WR}$, and after that it can be replaced by a new table. 

When the attack is detected, the filter measures the current query rates over the same windows. It then calculates the difference between the current rate $r_{cw_i}$ in the window $w_i$ and the rate expected by the model: $mean_{w_i} + 3 \times std_{w_i}$. We then calculate a smoothed, normalized deviance score $d_t$ at time $t$ as: $d_t = (d_{t-1}\times0.5) + 0.5\times\sum_i{\frac{r_{cw_i} - mean_{w_i} - 3*std_{w_i}}{std_{w_i}}}$. Those recursives whose deviance score exceeds threshold $t_{WR}$ will be identified as wild recursives.

We experimented with values for $L_{WR}$ between 10 minutes and 6 hours. While performance was relatively stable, lower values led to lower collateral damage, since they captured recent traffic trends. We experimented with uniformly distributed and exponentially distributed (powers of two) window sizes. Exponentially distributed windows led to lower mitigation delay, because they capture both aggressive and stealthy attackers. We also experimented with 1--9 windows. Higher number of windows had slightly higher collateral damage, but they significantly improved filter effectiveness, because they enabled us to identify sporadic attackers. Learned models become quickly outdated so we set $U_{WR} = L_{WR}$. We experimented with values for the threshold $t_{WR}$ from 0.1 to 16. Values higher than 0.5 minimized collateral damage.

\textit{Operational cost:} This filter is implemented by processing the traffic incoming to the DNS server offline. When the attack starts, the filter identifies wild recursives and inserts corresponding \texttt{ipset} rules to block their traffic.

\textbf{Response code filter (RC).}
	\label{sec:rcode} For some DNS servers, queries with missing names are rare.
For example, at Akamai only a small fraction of legitimate queries result in NXDomain~\cite{schomp2020akamai} replies, while attackers often query for random query names. 
We therefore considered a filter based on response codes that discards NXDomain responses.
Unfortunately, more than 60\% of \emph{root} DNS traffic involves non-existing TLDs.
Thus for root DNS traffic, a response code filter will have large collateral damage, and we do not currently include it in \system.

\textbf{Aggressive recursive filter (AR).}
\label{sec:agg}
This filter blocks the aggressive clients during an attack, starting with the client that sends the highest query rate and moving down. Filter adds addresses to the block-list until the query load reduces to acceptable levels. We evaluated this filter on our dataset. It performs well when attacks use non-spoofed traffic, but its performance is consistently worse than that of wild recursive filter. We thus do not include it in \system.

\begin{table*}
\centering
\vspace{0.05in}
\begin{tabular}{l|lllll|rr|rr|rr|rr|rr|rr}
\multirow{2}{*}{\textbf{PoP}} &
\multirow{2}{*}{\textbf{date}} & \textbf{start} & {\textbf{dur}} & \multirow{2}{*}{\textbf{ULQ}} &
 \multirow{2}{*}{\textbf{DNS}} &
 \multicolumn{2}{|c|}{\textbf{FQ}} & \multicolumn{2}{|c|}{\textbf{UR}} &
\multicolumn{2}{|c|}{\textbf{HC}} &
\multicolumn{2}{|c|}{\textbf{WR}} & \multicolumn{2}{|c|}{\textbf{\system$_F$}} &
\multicolumn{2}{|c}{\textbf{\system$_P$}} \\\cline{7-18}
&& (UTC) & (sec) &  & mon & \textbf{con} & \textbf{cd} & \textbf{con} & \textbf{cd} & \textbf{con} & \textbf{cd} & \textbf{con} & \textbf{cd} & \textbf{con} & \textbf{cd} & \textbf{con} & \textbf{cd}\\\hline
LAX & 2015-11-30 & 06:50 & 8,918* & 98 &  100 & \textbf{100} & \textbf{0} & \textbf{99.1} & \textbf{1.8} & 0.3 & 1.4 & 0 & 5.5 & \textbf{99.1} & \textbf{0.4} & 99.3 & 1.7\\
LAX & 2015-12-01 & 05:10 & 3,781* & 100 &  100 & \textbf{98.7} & \textbf{0} & \textbf{99.1} & \textbf{0} & 0.6 & 0 & 0 & 0 & \textbf{99.3} & \textbf{0} & \textbf{99.4} & \textbf{0}\\
LAX & 2016-06-25 & 22:18 & 2,436* & 52 &  99 & 0 & 0 & \textbf{100} & \textbf{0.1} & 0 & 0 & 0 & 0 & \textbf{100} & \textbf{0.1} & \textbf{100} & \textbf{0.1}\\
LAX & 2017-02-21 & 06:40 & 6,992* & 2 &  1 & \textbf{98.4} & \textbf{0} & 0.1 & 1.8 & 0.1 & 1.5 & 98.4 & 0 & \textbf{99} & \textbf{0} & \textbf{98.8} & \textbf{0}\\
LAX & 2017-03-06 & 04:43 & 19,835* & 6 & 5 & \textbf{98.8} & \textbf{0} & 0 & 1.1 & 0 & 0.4 & 91.6 & 1.5 & \textbf{100} & \textbf{0} & 92.3 & 1.5\\
LAX & 2017-04-25 & 09:54 & 10,414* & 3 & 4 & \textbf{98.3} & \textbf{0} & 0 & 1.1 & 0 & 0.7 & 94.9 & 2 & \textbf{99.1} & \textbf{0} & 95.1 & 2\\
ARI & 2019-09-07 & 06:45 & 80 & 0 & 5 & 0 & 0 & \textbf{93.3} & \textbf{0.6} & 0 & 0.8 & 0 & 0.1 & \textbf{93.7} & \textbf{0.6} & \textbf{93.1} & \textbf{0.6}\\
LAX  & 2019-09-07 & 06:45 & 80 & 23 & 5 & 0 & 0 & \textbf{100} & \textbf{0.9} & 0 & 0.2 & 0 & 0.2 & \textbf{100} & \textbf{0.9} & \textbf{100} & \textbf{0.9}\\
MIA & 2019-09-07 & 06:45 & 80 & 8 & 5 & 0 & 0 & \textbf{100} & \textbf{0.6} & 0 & 0 & 0 & 0.4 & \textbf{100} & \textbf{0.6} & \textbf{100} & \textbf{0.6}\\
SIN & 2020-02-13 & 08:05 & 206 & 14 & 2 & \textbf{100} & \textbf{0} & 0 & 0.3 & 4.8 & 0 & 38.5 & 0.5 & \textbf{100} & \textbf{0} & 97.5 & 0.8\\
ARI & 2020-10-24 & 02:55 & 445 & 67 & 7 & 0 & 0 & \textbf{100} & \textbf{1.3} & 0 & 0 & 0 & 0.8 & \textbf{100} & \textbf{1.3} & \textbf{100} & \textbf{1.3}\\
ARI & 2021-05-28 & 02:35 & 70 & 25 &  3 & 0 & 0 & \textbf{100} & \textbf{1.1} & 0 & 0 & 0 & 0.1 & \textbf{100} & \textbf{1.1} & \textbf{100} &\textbf{1.1}\\
IAD & 2021-05-28 & 02:35 & 70 & 63 & 3 & 0 & 0 & \textbf{100} & \textbf{0.4} & 0 & 0 & 2.7 & 0 & \textbf{100} & \textbf{0.5} & \textbf{100} & \textbf{0.5}\\
LAX & 2021-05-28 & 02:35 & 70 & 3 & 3 & 0 & 0 & \textbf{100} & \textbf{0.4} & 0 & 0 & 0 & 0 & \textbf{100} & \textbf{0.4} & \textbf{100} & \textbf{0.4}\\
MIA & 2021-05-28 & 02:35 & 70 & 2 & 3 & 0 & 0 & \textbf{100} & \textbf{1.5} & 0 & 0 & 0 & 0 & \textbf{100} & \textbf{1.7} & \textbf{100} & \textbf{1.7}\\
SIN & 2021-05-28 & 02:35 & 61 & 41 & 3 &  0 & 0 & \textbf{100} & \textbf{0} & 0 & 0 & 0 & 0 & \textbf{100} & \textbf{0} & \textbf{100} & \textbf{0}\\
\end{tabular} 
\caption{\system performance: comparing load control (con) and collateral damage (cd) for each possible filter and \system as a whole. We highlight results within 1\% of the best performance in bold. For long attacks (*) we simulate only the first 600 seconds.}
\label{tab:multi} 
\vspace{-0.27in}
\end{table*}

\subsection{Filter Selection and Synchronization}

In this section we discuss how filters are selected for deployment and why their learning periods have to be synchronized.

\textbf{Filter selection.}
Our goal was to design effective filter selection process, which minimizes collateral damage to legitimate traffic. Our pseudocode for filter selection is given in \autoref{fig:pseudocode}. At each time interval (e.g., one second), if the current query load ($CL$) on the server (queries per second) is higher than the acceptable load ($AL$), we
first select candidate filters. We continuously emulate operation of all filters, thus we produce for each filter an estimate of the amount of queries they would drop. Our candidate filters are those whose drop estimates are positive.
If among the candidate filters there are any that could reduce the load to $AL$, we will select the filter with the lowest estimated collateral damage (described in \autoref{sec:operation}) and deploy only this filter (function \textit{deploy\_single}). 

If no such filters exist, we will consider combinations of multiple filters (function \textit{deploy\_combo}). Not all combinations are valid, which greatly reduces complexity of this step. HC filter must be deployed after an UR filter, since HC is pass-through for addresses that do not exist in TTL-table. UR filter removes queries that spoof unknown recursives, thus guaranteeing that addresses of queries that pass will be present in TTL-table. FQ$_{t}$ could be deployed together with any other filter. FQ$_{s}$ and WR filters must be deployed after UR and HC, because they make per-source blocking decisions, and require reliable source identities. Since both FQ$_{t}$ and FQ$_{s}$ filter frequent query names, only one of them should be deployed. FQ$_{t}$ has zero collateral damage and is considered first. If it cannot be supported operationally (there are more than five query names, and thus there will be added processing delay), FQ$_{s}$ will be considered. 
In addition to considering filters in a specific order for deployment, we only consider filters that are \textit{effective} -- filter at least 5\% of excess traffic (function \textit{effective}). Deployment is finalized as soon as the filter combination can reduce the load below $AL$.

\textbf{Filter synchronization.} \label{sec:sync}
\system may engage one or multiple filters to mitigate an attack. When some filter combinations are engaged, it is important that their learning periods match, so that each filter has entries for the same recursives in their table. 
 Because we need a shorter learning period for wild recursive filter, than for the unknown recursive and hop-count filter, we learn parameters over $2$ hours, and then keep updating WR entries each $20$ minutes to keep them as recent as possible. 

\begin{figure}
    \centering
    \includegraphics[width=0.8\columnwidth]{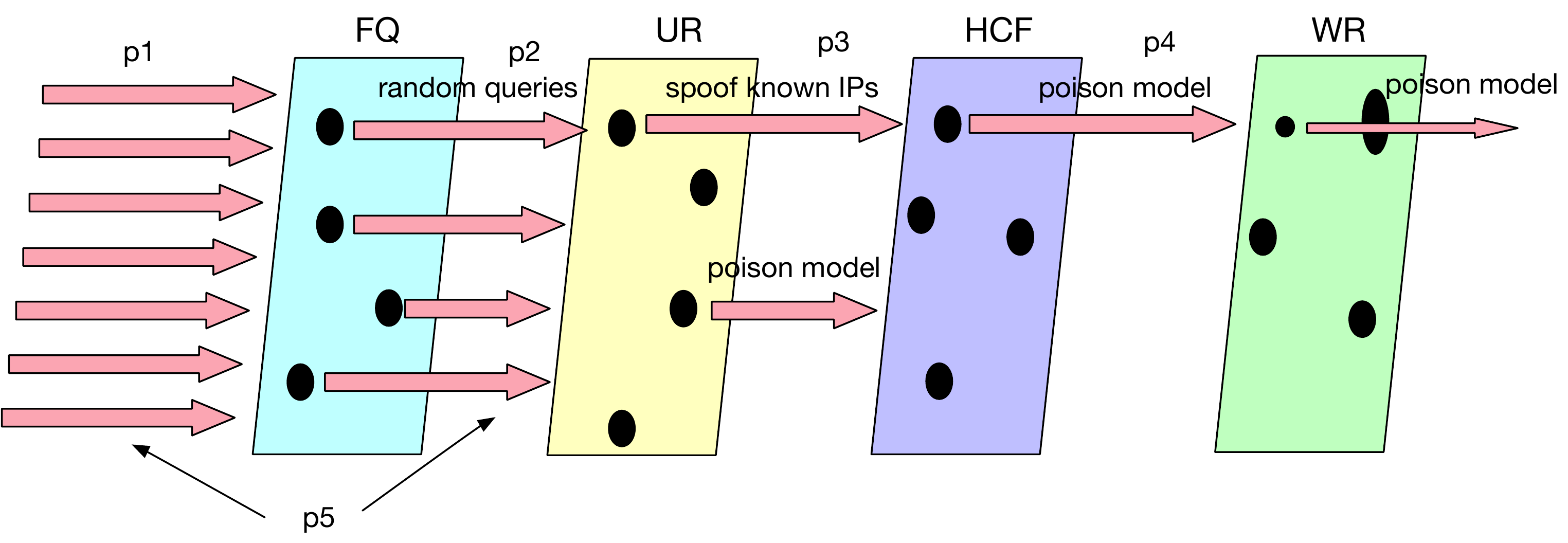}
    \caption{Swiss cheese model of defense}
    \label{fig:swiss}
        \vspace{-0.27in}
\end{figure}

\textbf{Sophisticated adversary.} 
Each of the filters we consider could be bypassed by a sophisticated adversary. We now discuss how their combination makes this challenging (\autoref{fig:swiss}).

FQ filter could be bypassed by the attacker sending random queries. UR filter could be bypassed by the attacker spoofing existing (known) recursives. UR, HC and WR filters could each be bypassed by poisoning the models during learning. One way to counter poisoning attacks could be to learn over longer time periods, from random traffic samples. While this works for UR and HC, whose data is fairly stable, it would greatly diminish effectiveness of WR filter, and it would complicate filter synchronization. Our approach is to handle poisoning attacks only at WR filter, and to rely on the Swiss cheese defense model (\autoref{fig:swiss}) to capture attackers that bypass one filter layer, but can be stopped at the other. Thus random queries may bypass FQ, but will be stopped at UR if they are from new sources, or at HCF if they are spoofed. At WR, queries sent by recursives at high rate (spoofed or not) can be detected and dropped. This leaves poisoning attacks at WR filter (thin red arrow at the top right of \autoref{fig:swiss}), where
 each bot poisons the rate model for itself by sending sporadic traffic during learning, with high fluctuations. This can lead the filter to model a large expected rate for the bot in each window, due to large standard deviation. To address this attack, we learn only when load on the server is low ($avg+stdev$). This forces the attacker to engage their bots very sporadically, which becomes an outlier and is excluded from the model. 

\section{Evaluation}
		\label{sec:eval}

We use datasets containing real DNS root traffic and attacks (\autoref{sec:data})
  to calculate success metrics (\autoref{sec:metrics})
  that characterize \system performance (\autoref{sec:perf}).

\subsection{Datasets}
\label{sec:data}
We use datasets collected at \broot, one of 13 root identifiers. 
These datasets are publicly available\ifisanon~\cite{dsnanonAnon}\else~\cite{attackdatasets} \fi
  in both pcap and text format.
The operators of \broot{}
  identify attacks based on unusual traffic rates and system load, as seen from operational monitoring.
Our evaluation uses ten diverse attack events spanning six years (see \autoref{tab:multi}).
During events in 2017 and later \broot{} employed anycast network
  with multiple points-of-presence (PoPs).
Some attacks affected only one PoP (e.g., 2020-02-13), while others targeted all PoPs (e.g., 2020-05-28).

We confirm that our selected events
  are DDoS attacks based on DNSmon observations shown in the ``DNSmon'' column
   \autoref{tab:multi}.
DNSmon reports the fraction of responses received by
  many (about 100) physically distributed probers,
which query each DNS root every 10 minutes.
In \autoref{tab:multi}, the first three attack events had a large impact,
  showing 99--100\% of unanswered queries,
  as publicly reported~\cite{moura2016anycast, root-news-nov, root-news-june}.
The other seven events had smaller impacts (1--7\% unanswered queries), because they were shorter (5 minutes and less)
  and sent at a lower rate,
  and because \broot's capacity had increased.
DNSmon reports reflect aggregate performance across all PoPs,
  so the percentage of unanswered queries at
  each PoP might be higher than measured by DNSmon. 
We include traces from all the PoPs in our analysis, and
simulate running of \system at each PoP. We use ground truth for attack start and stop times to start and stop \system's simulation, and use $f_{ACC}=2.5$. During attacks, query rate at the server increases more than 10-fold, so using $f_{ACC}=2.5$ is reasonable.



While attackers could generate any random traffic to port 53, attacks in our dataset had unique content or traffic signatures, which enabled us to establish \textit{ground truth} during evaluation. Attacks on 2015-11-30, 2015-12-01, 2017-02-21, 2017-03-06, 2017-04-25, and 2020-02-13 had used either several specific queries or a random prefix with a common, specific, suffix. Attack on 2016-06-25 was a TCP SYN flood. Attacks on 2019-09-07 and 2020-10-24 and 2021-05-28 sent malformed UDP traffic to port 53, which consumed resources at the server, but did not parse into legitimate query format. 

\textbf{Ethical considerations.}
Our analysis is performed on packet traces incoming to and outgoing from \broot{}.
Both source and destination IP addresses are anonymized using Crypto-PAn\ifisanon~\cite{xu2002prefix}.\else~\cite{xu2002prefix,dnsanon}.\fi ~Packet payloads are not anonymized, which allows us to establish ground truth in evaluation.
After ground truth is established,
  analysis is automated and we report only aggregate results.
These steps preserve resolver privacy.

\subsection{Metrics}
	\label{sec:metrics}


Our goal is to reduce load 
  on the DNS root server, by filtering malicious traffic, to allow serving more
  legitimate users when under duress.
We therefore consider two success metrics: (1)
  \emph{controlled load}, the percent of time when server load is at or below acceptable load due to defense's actions, ideally 100\%;
(2) \emph{collateral damage}, the percent of legitimate queries filtered, with an ideal of 0\%.

\subsection{\system Performance}
\label{sec:perf}

\autoref{tab:multi} shows \system's performance per each PoP affected by a given attack.
We show several defense configurations:
  first, each filter by itself (FQ, UR, HC, or WR), 
then the full \system with all four filters
  and a partial \system with only UR, HC, and WR filters.
Removing the FQ filter from the partial \system simulates 
 a smart adversary, which randomizes queries for each attack.

These experiments confirm that 
  \emph{no single defense does well in all attack cases}.
The FQ filter does very well in attacks that use similar queries, but has no effect otherwise.
The UR filter performs well in many attacks.
HC does not work well by itself, but enhances other filters.
Finally,
  WR does well in a few attacks, where some recursives, which are present prior to the attack, modify their behavior to become more aggressive.
This evaluation demonstrates that we need multiple filters to handle all attack events.

We further show that 
  \emph{the full \system automatically chooses the best filter or combination of filters
  for each attack}, always achieving 93\% or higher controlled load
  and at most 1.7\% collateral damage. 
\system selects the optimal filter combination in 1--3 seconds. 

Partial \system's performance (the right-most column) shows
how well it would handle \emph{an adversary that randomizes queries}.
\system controls load for most of the time (92.3\%--100\%),
  with low collateral damage (2\% or lower),
  with all filters selected in 3\,s or less.

 We compare collateral damage of \system with percentage of legitimate queries at the affected PoP that fail to receive a response during the original attack, without \system. We calculate this percentage from our datasets and show it in the fifth column (ULQ) of \autoref{tab:multi}. This is an internal measure of DoS impact and it can differ from the external measurements by DNSmon, because of several reasons. First, DNSmon averages measurements over 10 minutes and across all PoPs for a given root, while our internal-DoS measure is per PoP and it is averaged over the duration of the attack. For these reasons DNSmon will often underestimate attack impact, as is the case for many of our attacks. Second, if \broot{}'s incoming bandwidth were overloaded, DNSmon could measure higher loss rate than our internal-DoS measure. This is the case, for example, for 2019-09-07 attack. 
 
 Full \system's and partial \system's collateral damage is always lower than DNSmon (external) and ULQ (internal) measures. Thus \system improves legitimate traffic's handling during DoS attacks. \system is also effective, reducing resource consumption by controlling server load, 93--100\% of time, after a short initial delay of 1--3 seconds.  

\textbf{Legitimate flash crowds.} While three attacks in 2017 overloaded \broot, they involved a large number of recursives involved (around 50~K per event), large difference in rates per recursive, and did not spoof. Legitimate flash crowds would show similar patterns. In 2017 events, \system dropped only traffic that was causing the overload event, and only as much as to free server resources from overload. 

\textbf{Polymorphic attacks.}
In evaluation events \system changes defenses because the attacks change.
During 2015-11-30 attack there were periods where existing clients were spoofed with incremental TTL values, traversing the entire TTL value space.
Partial \system correctly switched from UR to UR+HC combo to handle these cases. During 2020-02-13 attack, single UR, HC and WR filters could not sufficiently reduce the load. Partial \system deployed all three filters, which managed to reduce the load.

\begin{figure}
    \centering
    \includegraphics[width=\columnwidth]{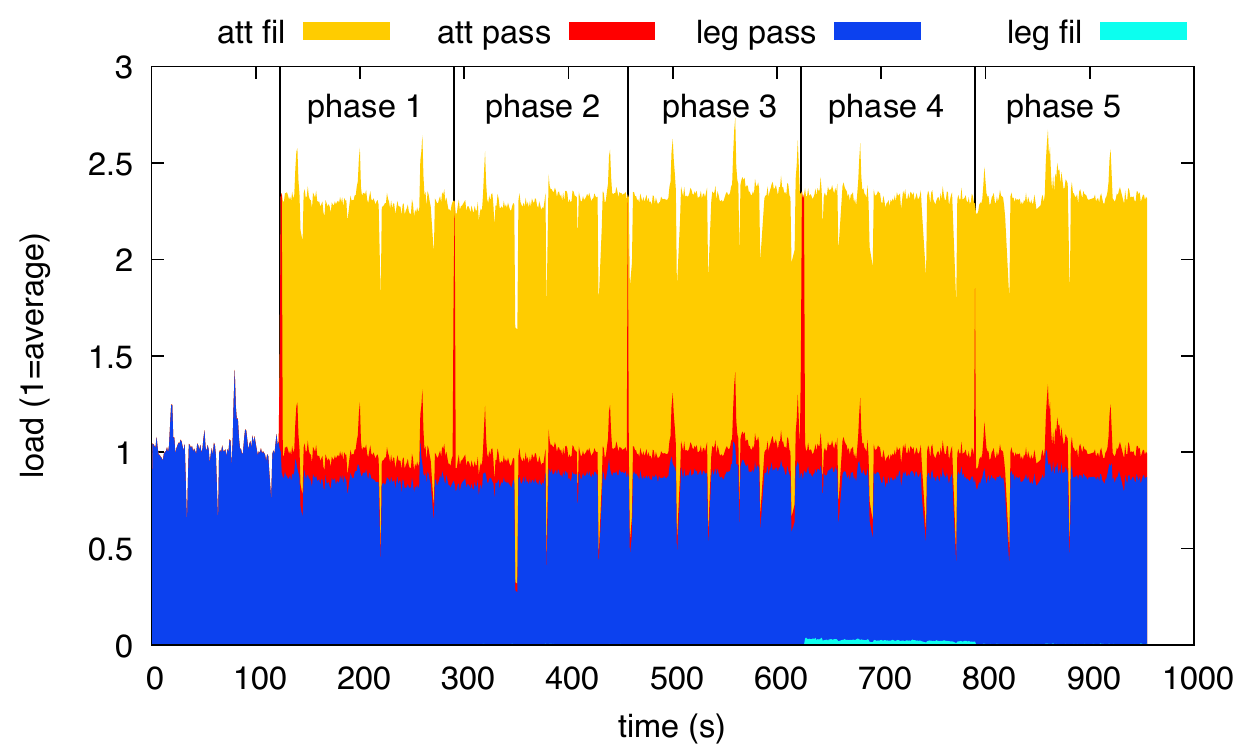}
    \caption{\system evaluation for a synthetic polymorphic attack.}
    \label{fig:var}
    \vspace{-0.27in}
\end{figure}

We demonstrate how \system can nimbly adjust filter selection by using an artificial polymorphic attack in \autoref{fig:var}.
We create a synthetic attack
  by mixing legitimate traffic from February 2017
  with five synthetic attacks, which correspond to p1--p5 labels in \autoref{fig:swiss}:
 (p1) a random-spoofed attack with a fixed query name,
  (p2) an attack with random query names,
  (p3) same as (p2) but also spoofs only known recursives using random TTL values,
  (p4) same as (p3) but spoofs with correct TTL values,
  (p5) same as (p1) but 90\% of queries are random and 10\% use a fixed query name.
We find that \system quickly converges to the best single filter for each attack strategy: FQ$_t$, UR, HC, WR and FQ$_s$, respectively.
\autoref{fig:var} shows passed and filtered legitimate and attack traffic for our synthetic attack---overall controlled load was 99.1\%, collateral damage was 0.7\%, and selection delay was 1--4\,s.

\section{Conclusion}

\iffalse
%
%
This paper provides the first in-depth design and evaluation 
  of an automated, layered approach to mitigate DDoS on DNS root. 
Evaluated on ten real-world DDoS attacks on \broot, \system quickly selects the best filter or filter combination from a library of filters, 
  and deploys it automatically. \system reduces server load to acceptable levels within seconds, with collateral damage under 2\%.
\system is adaptive to polymorphic attack events, which 
  change attack pattern during an ongoing attack event, and nimbly makes new filter selection in up to 4 seconds. It further has low operational cost, working offline to process incoming traffic at the server, and producing filtering rules, which can be implemented at no added processing delays using \texttt{ipset}. We release \system at $<$link removed for anonymity $>$ as open source.

\else
%
%
This paper provides the first in-depth design and evaluation 
  of an automated, layered approach to mitigate DDoS on DNS root. 
Evaluated on ten real-world DDoS attacks on \broot, \system quickly selects the best filter or filter combination from a library of filters, 
  and deploys it automatically. \system reduces server load to acceptable levels within seconds, with collateral damage under 2\%.
\system is adaptive to the polymorphic attack events, which 
  change attack pattern during an ongoing attack event, and nimbly makes new filter selection in up to 4 seconds. It further has low operational cost, working offline to process incoming traffic at the server, and producing filtering rules, which can be implemented at no added processing delays using \texttt{ipset}. We release \system as open source.
\fi

\ifisanon
\else
\section*{Acknowledgments}

This work is partially supported by the National Science Foundation (grant NSF
OAC-1739034) and DHS HSARPA Cyber Security Division (grant 
SHQDC-17-R-B0004-TTA.02-0006-I), in collaboration with NWO\@.


\fi
\flushleft

{\footnotesize
\bibliographystyle{abbrv}
\bibliography{ref}
}
\end{document}